\documentclass[prl,aps,twocolumn,superscriptaddress,showpacs]{revtex4-1}

\usepackage{graphicx}
\usepackage{epstopdf}                                       

\begin{document}                  

\title{(C$_{4}$H$_{12}$N$_{2}$)[CoCl$_{4}$] -tetrahedrally coordinated Co$^{2+}$ without the orbital degeneracy}

\author{C. Decaroli}
\affiliation{School of Physics and Astronomy, University of Edinburgh, Edinburgh EH9 3FD, UK}

\author{A. M. Arevalo-Lopez}
\affiliation{School of Chemistry, University of Edinburgh, Edinburgh, EH9 3JZ, UK}

\author{C. H. Woodall}
\affiliation{School of Chemistry, University of Edinburgh, Edinburgh, EH9 3JZ, UK}

\author{E. E. Rodriguez}
\affiliation{Department of Chemistry of Biochemistry, University of Maryland, College Park, MD, 20742, U.S.A.}

\author{J. P. Attfield}
\affiliation{School of Chemistry, University of Edinburgh, Edinburgh, EH9 3JZ, UK}

\author{S. F. Parker}
\affiliation{ISIS Facility, STFC Rutherford Appleton Lab, Chilton, Didcot,  OX11 0QX, UK}

\author{C. Stock}
\affiliation{School of Physics and Astronomy, University of Edinburgh, Edinburgh EH9 3FD, UK}

\date{\today}

\begin{abstract}

We report on the synthesis, crystal structure, and magnetic properties of a previously unreported Co$^{2+}$ $S={3 \over 2}$ compound, (C$_{4}$H$_{12}$N$_{2}$)[CoCl$_{4}$], based upon a tetrahedral crystalline environment.  The $S={3 \over 2}$ magnetic ground state of Co$^{2+}$, measured with magnetization, implies an absence of spin-orbit coupling and orbital degeneracy.  This contrasts with compounds based upon an octrahedral and even known tetrahedral Co$^{2+}$ (Ref. \onlinecite{Cotton83:1961}) based systems where a sizable spin-orbit coupling is measured.  The compound is characterized with single crystal x-ray diffraction, magnetic susceptibility, infrared, and ultraviolet/visible spectroscopy.  Magnetic susceptibility measurements find no magnetic ordering above 2 K. The results are also compared with the previously known monoclinic hydrated analogue.

\end{abstract}

\maketitle  

\section{Introduction}

Molecular magnets containing 3\textit{d} transition metal ions have played a central role in the creation of new materials and guiding theory owing to the interplay between orbital and spin magnetism and the ability to tune both degrees of freedom through the surrounding ligand field environment.\cite{Kugel46:92}  It is the aim of this work to find new quantum molecular magnets which will aid in the understanding of strongly correlated and quantum phenomena.

Co$^{2+}$ (d$^{7}$) compounds have provided an excellent starting point for the creation of model low dimensional magnets owing to the delicate interplay between orbital properties and the surrounding crystalline electric field environment.\cite{Schoonevel116:2012,Banci52:1982,Jaynes1995:34}  Due to the orbital degree of freedom, magnets based upon Co$^{2+}$ in an octahedral field tend to be excellent realizations of low spin magnets which display strong quantum fluctuations.~\cite{Abragram1986}  The basic orbital properties of Co$^{2+}$ are summarized in Figure 1 for both octahedral ($Dq>0$, where $10Dq\equiv \Delta$ is the splitting between the $t_{2g}$ and $e_{g}$ levels) and tetrahedral ($Dq<0$) coordination.~\cite{Balhausen1962,Orgel23:1955,Pappalardo35:1961}  Co$^{2+}$ in an octahedral environment has an orbital degree of freedom resulting in an orbital triplet ground state (with an effective orbital angular moment $\tilde{l}=1$) with a spin $S={3\over 2}$.  Spin-orbit coupling splits this twelve-fold degenerate state into a doubly degenerate ground state with an effective $j={1\over2}$ and two excited states with $j={3\over2}$ and ${5\over2}$.\cite{Cowley13:88,Sakurai167:68}  This doubly degenerate ground state has been exploited to study the magnetic response of low-spin chains in CoNb$_{2}$O$_{6}$ and CsCoX$_{3}$ (with X=Br and Cl) salts where low-energy fluctuations within the doublet $j={1\over2}$ manifold reside well below the $j={3\over2}$ level set by spin-orbit coupling.\cite{Coldea327:2010,Nagler27:1983,Cabrera90:2014}  These systems have been used to study the physics of the S=${1 \over 2}$ Ising chain.  

\begin{figure}
\includegraphics[width=8.7cm]{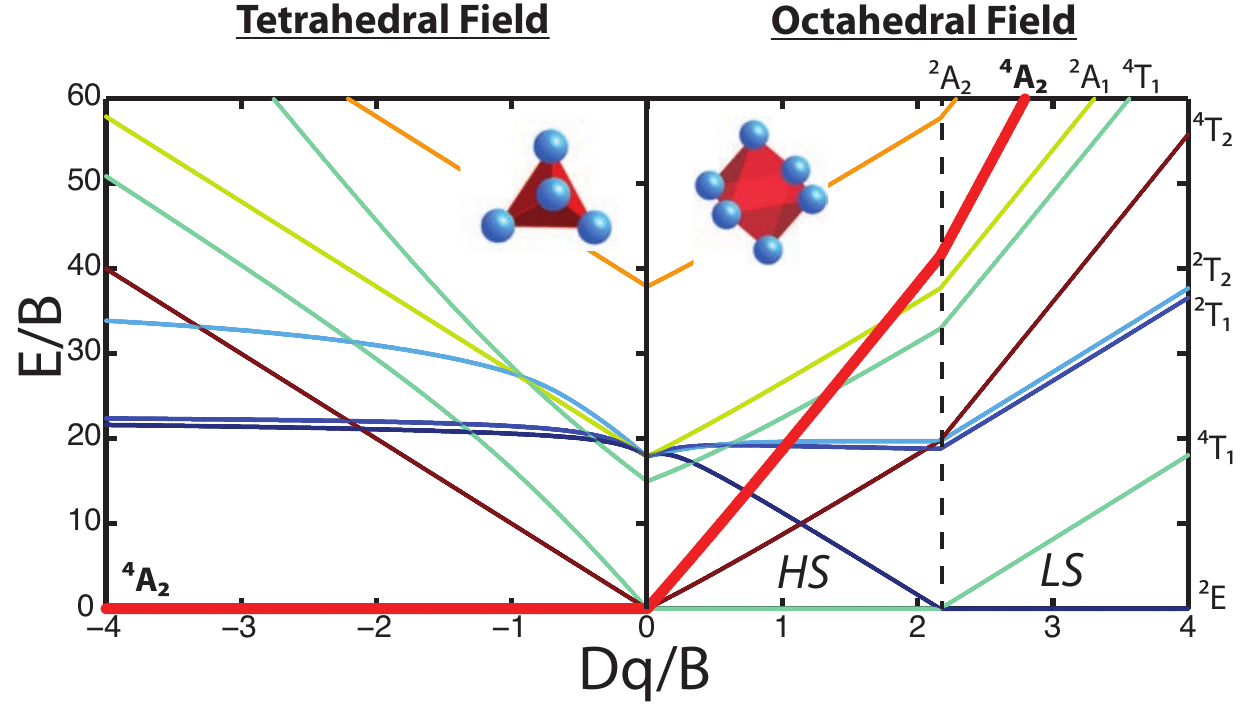}
\caption{\label{matlab} The Tanabe-Sugano diagram and orbital states for octahedral ($Dq>0$) and tetrahedral ($Dq<0$) crystalline electric field environments.  The orbital degeneracy in an octahedral crystalline field results in a spin transition from a $^{4}T_{1}$ (high spin - HS) state to a $^{2}E$ (low spin - LS) state for large values of $Dq$.  For negative values of $Dq$ (tetrahedral environment), the orbital triplet is replaced by a singlet state ($^{4}A_{2}$) with a  $S={3 \over 2}$. However, real systems display a coupling between this singlet ground state and the excited orbital triplet resulting in a ground state orbital degeneracy.  The diagram was calculated taking the Racah parameters $B$ and $C$ to be the free ion values of  0.12 eV and 0.56 eV respectively (see Ref. \onlinecite{Abragram1986} for discussion and further definitions).  This work is centred around the goal of discovering new $S={3\over 2}$ Co$^{2+}$ magnets based on a tetrahedral crystal field ($Dq<0$, left side of the plot).}
\end{figure}

It is interesting to search for $S={3\over2}$ low dimensional magnets given that these systems are not obviously classical or quantum mechanical (see for example, Ref. \onlinecite{Igarashi46:92} which discusses the relation between spin size and quantum corrections).   As reviewed in Figure 1, Co$^{2+}$ in a tretrahedral environment possesses a spin state of $S={3\over2}$ and this can be modelled by reversing the sign of the crystal field splitting.  In a tetrahedral environment, the orbital degeneracy and complications due to spin-orbit coupling present in the $^{4}T_{1}$ octahedral state are removed and replaced by an orbital singlet $^{4}A_{2}$ state with $S={3 \over 2}$.  Despite this expectation based upon crystal field arguments, Co$^{2+}$ in a tetrahedral environment typically does display an orbital degree of freedom~\cite{Cotton1999,Horrocks76:98,Romerosa16:2003} complicating the system and the ability to apply such model magnets to test theories described above.  

In this report, we discuss the magnetic and structural properties of Co$^{2+}$ in a series of piperazine based molecular magnets.  We will show that the magnetic ground state of these systems lack a strong measurable orbital degree of freedom making them candidate model examples of $S={3\over 2}$ magnets.  This paper is divided into five sections including this introduction.  We first note the solution based synthesis of the previously unreported Co$^{2+}$ (C$_{4}$H$_{12}$N$_{2}$)[CoCl$_{4}$] compound and compare this with the known growth of the hydrated version.  We then present the crystal structure and magnetic susceptibility illustrating the underlying $S={3 \over 2}$ nature of this compound.  We finally present spectroscopic data measuring the orbital transitions and compare these with other Co$^{2+}$ tetrahedrally based compounds.

\section{Synthesis}

Motivated by piperazine based Cu$^{2+}$ quantum magnets, we investigated similar Co$^{2+}$ based compounds.~\cite{Stone95:2006}  Piperazinium Tetrachlorocobaltate Monohydrate (C$_{4}$H$_{12}$N$_{2}^{2+}\cdot$CoCl$_{4}^{2-}\cdot$H$_{2}$O - denoted as $PTCM$, for the remainder of this paper) was originally reported by Tran Qui and Palacios and is monoclinic (space group P2$_{1}$/a) with $a$=14.017 \AA, $b$=12.706 \AA, $c$=6.559 \AA, and $\beta$=87.21$^{\circ}$.\cite{Qiu46:90}  The structure consists of tetrahedrally coordinated Co$^{2+}$ and piperazinium layers coupled only through hydrogen bonding.  The Co$^{2+}$-Co$^{2+}$ distances are large (the drawn bond lengths labelled $J$ in Figure 2 $b)$ are $\sim$ 6-7 \AA), particularly in comparison to other Co$^{2+}$ materials such as CoO  where the Co$^{2+}$-Co$^{2+}$ distances are $\sim$ 3-4 \AA.\cite{Greenwald53:6}  Because of the large distances the magnetic properties were not characterized and the tetrahedra considered as isolated.   Based on evaporation out of solution, we were able to synthesize either $PTCM$ or the previously unreported linear chain variant (C$_{4}$H$_{12}$N$_{2}$)[CoCl$_{4}$] (termed $PTC$ during this report).  CoCl$_{2}$ and piperazine (C$_{4}$H$_{10}$N$_{2}$) were mixed in a solution in a 1:4 molar ratio. Piperazinium is the conjugate acid of the basic piperazine building block and is formed upon protonation in HCl. Therefore, both compounds were dissolved separately in concentrated (37\%) hydrochloric acid and then mixed in a single solution.  

\begin{figure}
\includegraphics[width=8.7cm]{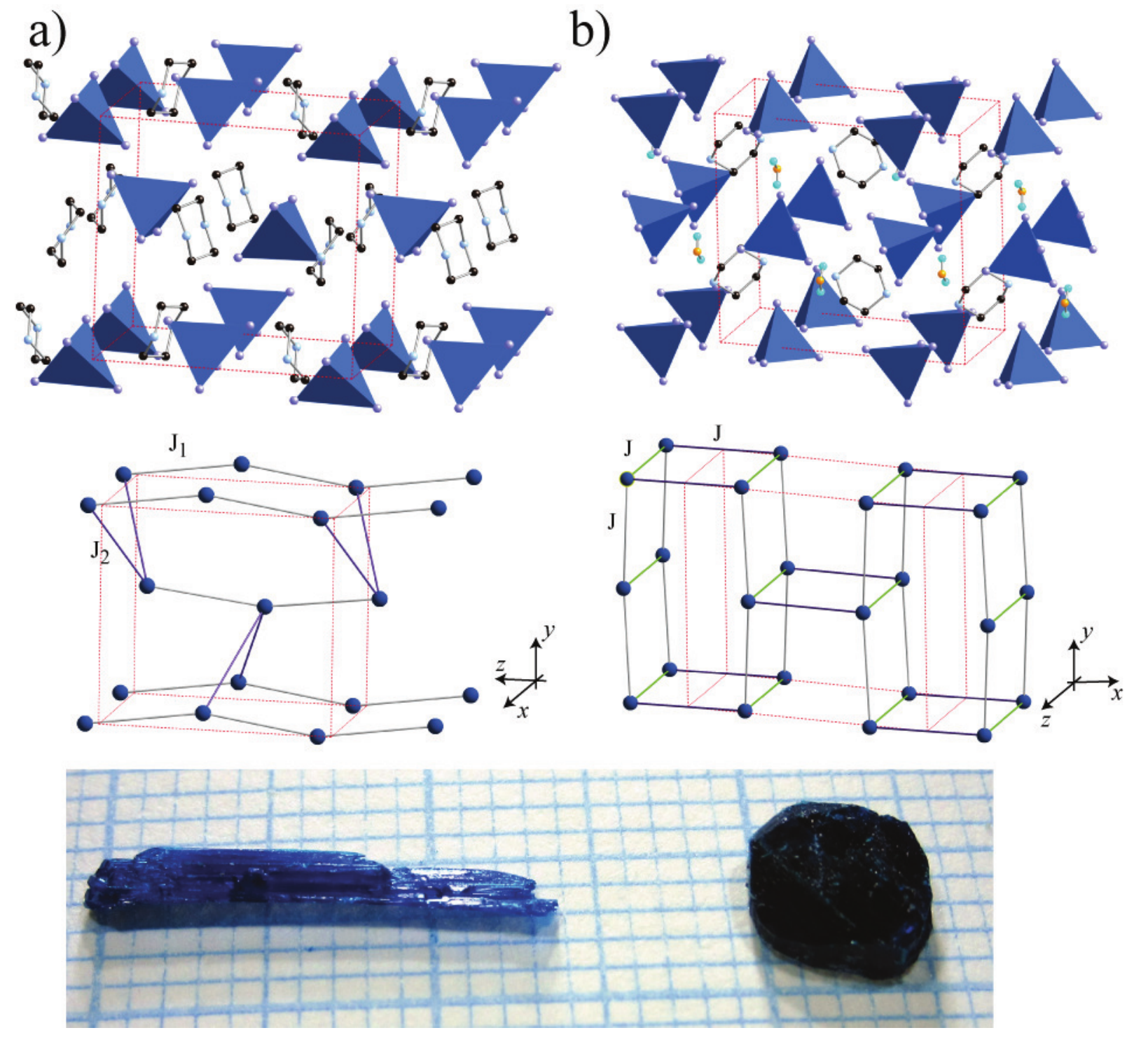}
\caption{\label{structure} $a)$ displays the structure of (C$_{4}$H$_{12}$N$_{2}$)[CoCl$_{4}$] -``$PTC$" with inter ($J_{1}$) and intra ($J_{2}$) exchange pathways highlighted.  $b)$ compares this to the structure of -``$PTCM$".  The lower panel shows typical crystal morphologies on millimetre scaled paper.}
\end{figure}

We initially followed the synthesis procedure for Cu$^{2+}$ quantum magnets based on piperazine and found two classes of molecules depending on the synthesis procedure.  The evaporation temperature and hydrated nature of the CoCl$_{2}$ were found to be key to whether crystals of $PTCM$ or $PTC$ were produced.  Rapid evaporation (temperatures between 60 $^{\circ}$ and 200 $^{\circ}$) formed $PTC$ needle crystals.  Typical lengths were 3-5 mm, however crystals up to a few centimetres in length were also synthesized (Figure 2 $a)$). Anhydrous CoCl$_{2}$ was found to always produce $PTC$ crystals while hydrated versions were found to produce $PTCM$ ($PTC$) from low (high) temperature evaporation.  The needle-like nature of $PTC$ contrasts with the cuboid nature of $PTCM$ formed at lower temperatures (Figure 2 $b$). 

\section{Single crystal x-ray diffraction}

Single crystals of both $PTCM$ and $PTC$ were isolated and their structures were studied via single crystal x-ray diffraction as summarized in Tables \ref{table_frac} and \ref{table_cryst}.  Experimental information is provided in Table \ref{table_xray}.    The structure of $PTCM$ was verified, using a similar experimental configuration to that listed in Table \ref{table_xray}, to be the same as reported previously.~\cite{Qiu46:90} $PTC$ was found to be isostructural with the Zn analogue reported by Sutherland \textit{et al}.\cite{Sutherland65:2009} The spacing between the tetrahedra in the lattice environment shows an anisotropy in one direction which has led us to investigate whether this is a candidate $S={3 \over 2}$ chain.  The distance between the tetrahedra for the two nearest neighbours is given in Table \ref{table_cryst},  whereas the next largest directions are 8.32 \AA\ and 8.12 \AA.   The difference between $J_{1}$ and $J_{2}$ defines the growth direction in Figure 2. This contrasts with $PTCM$ which has a definitive three dimensional structure. 

\begin{table}
\caption{Single crystal X-ray diffraction data for PTC taken with a Bruker APEX-II CCD diffractometer at room temperature.}
\centering
\begin{tabular} {c c}
\hline\hline	
Space Group			& 	P2$_{1}$2$_{1}$2$_{1}$ \\
\textit{a} (\AA)				&	8.2876(6)  \\
\textit{b} (\AA)				&	11.1324(9)  \\
\textit{c} (\AA)				&	11.9121(9)  \\
Crystal System		&	orthorhombic \\
Volume (\AA$^{3}$)			&	1099.02(15) \\
Z				&	4 \\
Formula weight  	&	288.89 \\
Calculated density (mg/mm$^{3}$) &	1.746 \\	 
$\lambda$ Mo K$\alpha$ (\AA) &	0.71073 \\
Monochromator 		 &	graphite \\
no. of reflections collected	& 	40260 \\
no. of independent reflections		&	4190 \\	
Absorption coefficient (mm$^{-1}$)	&	2.480 \\
F(000)					&	580.0 \\
R$_{1}$, wR$_{2}$ (\%) &	4.35, 7.40 \\	
$R_{1}=\sum||F_{0}|-|F_{c}||/\sum |F_{0}|$ &	\\
$wR_{2}=\left( \sum \left [w (F_{0}^{2}-F_{c}^{2})^{2} \right ]/[w(F_{0}^{2})^{2}] \right)^{1/2}$ &	\\
\hline
\label{table_xray}
\end{tabular}
\end{table}

\begin{table}
\caption{The fractional coordinates ($\times$ 10$^{4}$) for the Co and Cl positions in PTC.}
\centering
\begin{tabular} {c c c c c}
\hline\hline	
\textit{Atom}		& 	\textit{x}	&	\textit{y} 	& 	\textit{z} 	& \textit{U(eq)}\\
\hline 
Co1	&	2285.4(3)		&	358.0(2)		&	5918.4(2)  	& 	32.79(7)	\\
Cl1	&	4362.5(6)		&	-977.8(5)		&	5941.7(5)  	& 	39.06(11)	\\
Cl2	&	2378.7(6)		&	1419.6(4)		&	7541.6(4)  	& 	36.25(11)	\\
Cl3	&	2513.9(7)		&	1501.7(5)		&	4360.4(4)  	& 	43.21(13)	\\
Cl4	&	-136.2(6)		&	-658.4(5)		&	5981.7(5)  	& 	38.67(11)	\\
\hline
\label{table_frac}
\end{tabular}
\end{table}

\begin{table}
\caption{Two shortest Co$^{2+}$-Co$^{2+}$ ($J_{1,2}$), neighbour distances, and angles for the Co$^{2+}$ and Cl$^{-}$ positions for (C$_{4}$H$_{12}$N$_{2}$)[CoCl$_{4}$] (termed $PTC$ in the main text)  refined in orthorhombic P$2_{1}2_{1}2_{1}$ space group with cell parameters $a$=8.2876(6)\AA, $b$=11.1324(9)\AA, $c$=11.9121(9)\AA.}
\centering
\begin{tabular} {c c c c}
\hline\hline	
Co-Co($J1$)	& 6.043(1) \AA	&	Co-Co($J2$) & 7.743(1)\AA \\
\hline 
Co1-Cl1	&	2.2749(6) \AA		&	Co1-Cl2	&	2.2674(6)\AA \\
Co1-Cl3	&	2.2586(6) \AA		&	Co1-Cl4	&	2.3052(6)\AA \\
\hline\hline 
Cl1-Co1-Cl4	&	109.73(2)$^{\circ}$		&	Cl2-Co1-Cl1	&	107.73(2)$^{\circ}$ \\
Cl2-Co1-Cl4	&	104.93(2)$^{\circ}$		&	Cl3-Co1-Cl1	&	108.36(2)$^{\circ}$ \\
Cl3-Co1-Cl2	&	113.83(2)$^{\circ}$		&	Cl3-Co1-Cl4	&	112.12(2)$^{\circ}$ \\
\hline
\label{table_cryst}
\end{tabular}
\end{table}

\section{Magnetic susceptibility}

Magnetic susceptibility measurements were performed on a Quantum Design MPMS system.  Figure 3 $a)$ illustrates a plot of the magnetization as a function of magnetic field along directions parallel to the chain axis and perpendicular.  The high field limit tends towards $\sim$ 3 $\mu_{B}$, consistent with expectations based upon Co$^{2+}$ in a tetrahedral crystal field environment with $S={3\over2}$.  Panel $b)$ illustrates plots of the susceptibility and inverse for $PTC$ and $PTCM$, respectively, showing no sharp anomaly indicative of magnetic order above 2 K.  A Curie fit ($\chi \propto {1\over T}$) for $PTC$ gives a moment of 3.7 $\mu_{B}$, in close agreement to the expected result for $S={3\over2}$ of 3.9 $\mu_{B}$.  The moment is consistent with weak or negligible mixing with the excited $^{4}T_{2}$ (Fig. 1) state and non-measurable spin orbit coupling ($\lambda$) which would be reflected by an enhanced effective moment of $\mu=3.89-1.559 {\lambda \over |Dq|}$ (with $\lambda$ negative).~\cite{Cotton83:1961,Holm32:1960}  The lack of substantial spin orbit coupling is different from previously studied tetrahalo complexes.~\cite{Holm59:31} 

\begin{figure}
\includegraphics[width=8.7cm]{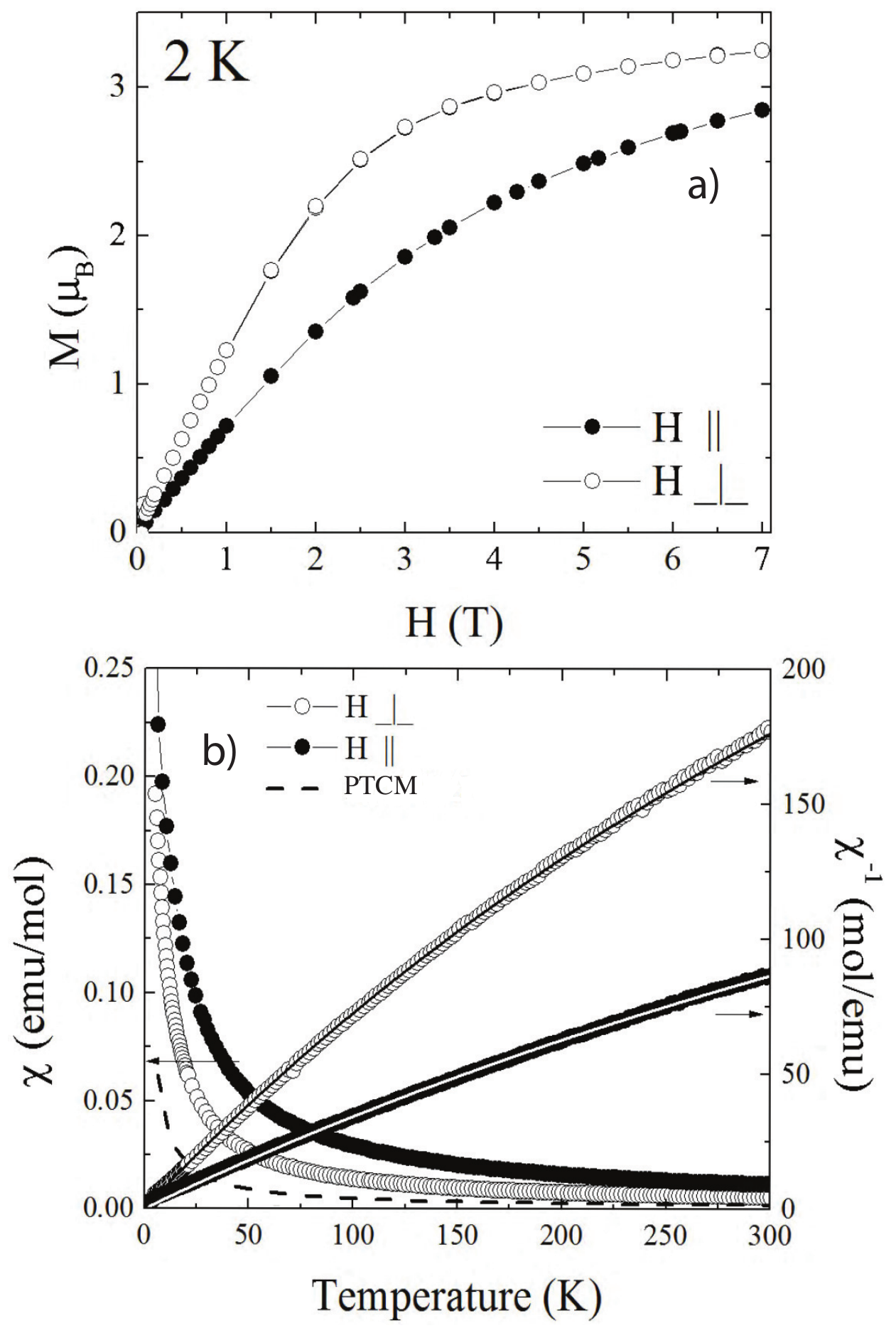}
\caption{\label{squid} $a)$ plots the magnetization of PTC as a function of applied field parallel and perpendicular to the chain axis.  The dashed lines are fits to a Brillouin function to extract a check for the $g$ factors derived from the temperature dependence.  The curves tend to $\sim$ 3 $\mu_{B}$ as expected given the crystalline electric field environment.  As confirmed by the fits, the saturation field is larger than the 7T field range probed.  $b)$ shows the susceptibility and inverse susceptibility for both $PTC$ and $PTCM$.  The solid lines are fits to a Curie type susceptibility ($\propto {1\over T} + A$).}
\end{figure}

The $g$ factors were derived from fits to the susceptibility (Fig. 3) to be $g_{||}$=2.49 and $g_{\perp}$=1.61.~\cite{Kuzian74:2006,Macfarlane47:1967}  The powder average, $\tilde{g}\equiv {1\over 3}g_{||} +{2 \over 3}g_{\perp}$=1.90, is in agreement with the expected value of 2 and confirms the lack of any mixing and spin-orbit coupling.  The ratio $g_{||}/g_{\perp}$=1.6 is close to the value of 1.8 derived from $M$ vs $H$ data (Fig. 3 $a)$ using Brillouin functions.   The anisotropic ratios of $g$ may reflect the slight crystallographic distortion of the tetrahedron.  As seen in Fig. 3, the susceptibility is described by a Curie term and a temperature independent term $\chi \propto {1\over T} +A$. The results were consistent with zero or negligible Curie Weiss temperature indicating very weak interactions, analogous to Cs$_{3}$CoCl$_{5}$ and Cs$_{3}$CoBr$_{5}$, with tetrahedrally coordinated Co$^{2+}$ with similarly large Co$^{2+}$-Co$^{2+}$ distances.\cite{Stapele66:1966} 

\section{UV-Vis and IR spectroscopy}

\begin{figure}
\includegraphics[width=8.7cm]{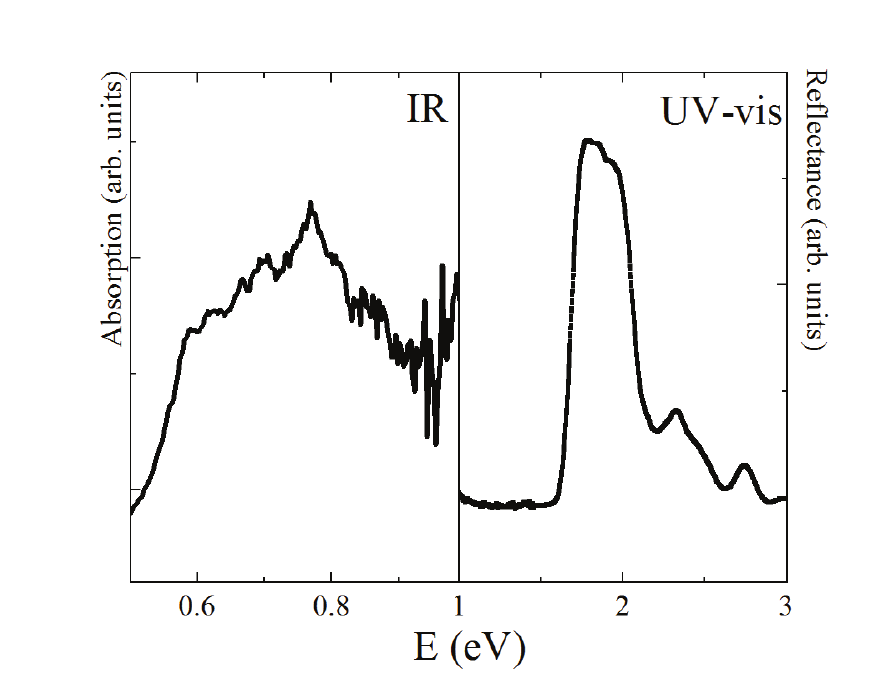}
\caption{\label{uv_plus_IR}  The IR and UV-Vis spectra for PTC illustrating electronic transition at $\sim$ 0.7 eV and at $\sim$ 1.8 eV.  Following previous studies based upon Co$^{2+}$ in a tetrahedral environment, we assign both of these transitions as excitations from the ground $^{4}A_{2}$ orbital singlet to the excited $^{4}T_{1}$ orbital levels.}
\end{figure}

Given the lack of any spin orbit coupling, in contrast to previous tetrahedrally coordinated Co$^{2+}$ complexes, we investigated the crystal field excitations to determine the orbital levels and compare them with previous Co$^{2+}$ compounds.  To extract the crystal field splitting and determine the energy scale of the orbital transitions for comparison with previous work, we performed both UV-Vis and IR spectroscopy.  UV-vis data were recorded in solution using a JASCO V-670 Series spectrophotometer.  Infrared spectra (400 - 8000 cm$^{-1}$) were recorded with a Bruker Vertex 70 FTIR spectrometer (TGS detector, 4 cm$^{-1}$ resolution, 64 scans) using attenuated total internal reflection (ATR) by a Bruker Platinum ATR accessory with a diamond ATR element. 

The results for both experiments are illustrated in Figure 4.  Two broad excitations are observed at $\sim$ 0.7 eV and 1.8 eV.   Following the systematic work by Cotton \textit{et al.} ~\cite{Cotton83:1961}, we assign these transitions to excitations from the $^{4}A_{2}$ orbital singlet to excited $^{4}T_{1}$ orbital triplets.  Based on the energy positions and the Tanabe-Sugano diagram (Fig. 1), we estimate that the crystal field splitting $Dq/B \sim 0.3$ implying $\Delta\equiv=10Dq\sim$ 350 meV ($\sim$ 2800 cm$^{-1}$).  It is interesting to note that the crystal field splitting is significantly less than octahedral variants such as NiO and CoO where $10Dq\sim$ 1 eV.~\cite{Kim84:11,Cowley13:88,Haverkort99:2007,Larson99:2007,Schoonevel116:2012}  The excitations are nearly identical (though slightly lower in energy) to those presented in the chlorine variants~\cite{Cotton83:1961} where mixing between orbital states was implicated as the origin of a spin orbit coupling resulting in magnetic moments $\sim$ 4 $\mu_{B}$ being observed.  Similar effects can be seen in other 4-coordinated cobalt compounds.~\cite{Wilson4:1985,Dong11:50}  Contrasting to these compounds, (C$_{4}$H$_{12}$N$_{2}$)[CoCl$_{4}$]  is an ideal example of an ionic material where no orbital mixing is present and the magnetic ground state is in a $S={3\over2}$ state with no spin orbit coupling.    

\section{Summary}

We have presented data on (C$_{4}$H$_{12}$N$_{2}$)[CoCl$_{4}$] - a $S={3\over 2}$ weakly coupled linear chain based upon Co$^{2+}$ in a tetrahedral environment.  The compound displays no strong spin orbit coupling and accommodates a $S={3\over2}$ ground state.   While, as noted above, most systems with Co$^{2+}$ in tetrahedra coordination display strong orbital effects as evidenced through large magnetic moments, there are some other counterexamples.  Most notably, Co$^{2+}$ in a square planar environment results in a low-spin configuration.~\cite{Everett1965:87} Divalent Co$^{2+}$ complexes supported by the [PhB(CH$_{2}$PPh$_{2}$)$_{3}$]$^{-}$ ligand also display small magnetic moments and hence small spin orbit effects.~\cite{Jenkins2002:124}    These complexes, similar to the compounds described here, were slightly distorted indicating the importance of a distortion away from a perfect tetrahedron to accommodate an $S={3\over2}$ ground state.  

In summary, $PTCM$ and $PTC$ crystals were both synthesized using fast and slow evaporation techniques respectively. Both crystals have magnetic properties determined by the cobalt ion Co$^{2+}$ with a d$^{7}$ electronic configuration in a tetrahedral environment. The resulting ligand environment forces the Co$^{2+}$ ion into an orbital singlet state removing the effects of orbital degeneracy and spin-orbit coupling present in the octahedral counterpart.  Susceptibility measurements find no observable magnetic order above 2 K in either $PTC$ or $PTCM$ and are consistent with weak exchange in both systems.  It would be interesting to pursue magnetic studies of similar systems which show significant exchange coupling as excellent realizations of marginal model quantum magnets. These compounds might prove to be useful model systems for testing theoretical predictions of quantum magnetism, $e. g.$ measuring quantum fluctuations directly with neutron scattering.  

We are grateful for funding from the Royal Society of Edinburgh, the Royal Society of London, and the Carnegie Trust for the Universities of Scotland.


%

\end{document}